\documentclass[aps,prl,twocolumn,groupedaddress,amsmath,amssymb,showpacs, floatfix]{revtex4-1}
 \usepackage{graphicx,graphics,color}
\usepackage{amsmath}
\usepackage{amssymb}
\usepackage{amsfonts}
\usepackage{amsfonts}
\usepackage{epstopdf}
\usepackage{bm}
\usepackage{times,xspace}
\usepackage{color}
\usepackage[utf8]{inputenc}
\usepackage{mathrsfs}
\usepackage{braket}
\DeclareMathOperator{\Tr}{Tr}

\DeclareFontFamily{U}{min}{}
\DeclareFontShape{U}{min}{m}{n}{<-> udmj30}{}

\DeclareMathOperator{\sinc}{sinc}

\begin{document}

\title{Non-hermitian time evolution: from static to parametric instability}

\author{Aleksi Bossart}
\author{Romain Fleury}
\affiliation{Laboratory of Wave Engineering, École Polytechnique Fédérale de Lausanne, 1015 Lausanne, Switzerland}


\begin{abstract}

Eigenmode coalescence imparts remarkable properties to non-hermitian time evolution, culminating in a purely non-hermitian spectral degeneracy known as an exceptional point (EP). Here, we revisit time evolution at the EP and classify two-level non-hermitian Hamiltonians in terms of the M\"obius group. We then leverage that classification to study dynamical EP encircling, by applying it to periodically-modulated (Floquet) Hamiltonians. This reveals that Floquet non-hermitian systems exhibit rich physics whose complexity is not captured by an EP-encircling rule. For example, Floquet EPs can occur without encircling and vice-versa. Instead, we show that the elaborate interplay between non-hermitian and modulation instabilities is better understood through the lens of parametric resonance.

\end{abstract}
\maketitle


\section{Introduction}

Non-hermitian Hamiltonians describe active systems, such as paraxial waveguides or coupled resonators with gain and loss \cite{miri_exceptional_2019}. Such traditional non-Hermitian systems are usually time-invariant, and exhibit standard instabilities originating from gain of static nature. More interestingly, the absence of a spectral theorem for these Hamiltonians gives rise to unconventional behaviour. The most striking example is a type of purely non-hermitian spectral degeneracy, in which the eigenvectors collapse as well: the so-called exceptional point (EP) \cite{Pancharatnam_1955,kato_perturbation_1995,wiersig_asymmetric_2008,heiss_time_2010,berry_half-century_2017,chen_revealing_2020,miri_exceptional_2019}. These degeneracies have a long history, dating back to studies of non-diagonalisable dielectric tensors by Pancharatnam \cite{Pancharatnam_1955} and mathematical studies of linear operators \cite{kato_perturbation_1995}, with a turning point marked by the advent of  pseudo-hermitian quantum theory \cite{bender_real_1998,mostafazadeh_pseudo-hermiticity_2002-1,mostafazadeh_pseudo-hermiticity_2002}. Both theoretical and experimental studies have shown that EPs allow for superior sensors \cite{wiersig_prospects_2020,dong_sensitive_2019,chen_exceptional_2017,liu_metrology_2016}, while recent theoretical studies started delineating the limitations of such EP-based sensing \cite{langbein_no_2018,lau_fundamental_2018}. Another promising application is that of EP-based polarizers, which have been explored in several studies \cite{doppler_dynamically_2016,hassan_dynamically_2017,zhang_dynamically_2018,zhang_dynamically_2019}.
 
The Riemann surfaces associated to EPs have also elicited much interest, both for topological reasons and as a way to circumvent the experimental difficulty of operating exactly at an EP, by encircling it instead \cite{dembowski_encircling_2004,mailybaev_geometric_2005,uzdin_observability_2011}. This process has since been realized in many experimental settings \cite{doppler_dynamically_2016,xu_topological_2016,yoon_time-asymmetric_2018,zhang_dynamically_2018,fernandez-alcazar_robust_2020}. In adiabatic EP encircling, the behavior of the system is well-understood with respect to the properties of the instantaneous Hamiltonian over the modulation trajectory. This is, however, not the case for general non-adiabatic periodic modulations, which unleash a whole new class of Floquet non-hermitian systems \cite{telnov_ab_2005,longhi_absence_2013,longhi_non-hermitian_2017,lee_pt-symmetric_2015,chitsazi_experimental_2017,li_observation_2019,koutserimpas_electromagnetic_2018} that may exhibit gain of parametric nature \cite{chitsazi_experimental_2017,li_observation_2019,koutserimpas_electromagnetic_2018}. While deviations from the adiabatic theorem have been studied in the case of particular modulation schemes, such as the circular trajectory \cite{berry_slow_2011,hassan_dynamically_2017,deng_exact_2019}, the role of exceptional point encircling and the importance of the modulation details (speed, strength, center) is still not well understood. A general theory and classification of static and Floquet non-Hermitian time-evolution is still missing and crucial to understand the interplay between the static non-Hermitian properties of the unmodulated system and periodic dynamic modulation.

In this paper, we present a classification of non-hermitian time evolution and use it to unveil the relationship between non-hermitian and parametric instabilities, in particular in relation with EP encircling. In the first section, we focus on time-independent non-Hermitian systems and explain how time evolution works in the absence of a full set of eigenmodes (i.e. at the EP). This allows us to introduce a general classification of time-evolution based on the Möbius group, and interpret it in terms of the dynamics of the state's polarisation. We show that even in generic cases, any initial condition converges to a fixed point in polarisation. In the second part of the paper, we apply our classification to periodically-modulated Hamiltonians, in order to understand the long-time dynamics of Floquet systems. We provide several examples to demonstrate that Floquet EPs can arise for time modulations that are neither close to nor encircle the static EP. Finally, we use our M\"obius group classification to unveil the origin of the different Floquet non-Hermitian classes, by connecting them to the phenomenon of parametric resonance.

\section{Classification of non-Hermitian dynamics}

In this section, we propose a general classification of non-Hermitian Hamiltonians. We begin with a short reminder of time evolution in hermitian quantum mechanics, to bring contrast with the non-hermitian situation. In the hermitian case, the Schrödinger equation

\begin{equation}
    i \hbar\frac{\partial}{\partial t}\ket{\psi(t)} = \hat{H}\ket{\psi(t)}
    \label{eq:schroedinger}
\end{equation}

can in principle be solved by putting the Hamiltonian $\hat{H}$ in diagonal form, thus effectively decoupling the evolution of all eigenstates $\ket{\boldsymbol{\phi}_k}$. They all follow a simple oscillating time evolution, 

\begin{equation}
    \ket{\boldsymbol{\phi}_k(t)}= e^{\frac{i E_k}{\hbar} t}\ket{\boldsymbol{\phi}_k(0)}
    \label{eq:HermSol}
\end{equation}

The time evolution for arbitrary states is thus easily determined by expressing them in the eigenstate basis of $H$, and then evolving each component according to Eq.\ref{eq:HermSol}. This is always possible in principle, since the spectral theorem guarantees that hermitian matrices are diagonalisable. In the following, we lift this hermiticity restriction and derive the consequences for time evolution, starting with the extreme case of an EP.


\subsection{Exceptional points}

 We consider a general $2 \times 2$ Hamiltonian,

\begin{equation}
	H := \begin{pmatrix}
		a & b\\
		c & d
	\end{pmatrix},
\end{equation} 

with unrestricted complex entries. Since we are interested in EPs, let us see when the eigenvalues coalesce. To this end, we compute the characteristic polynomial and complete the square, 

\begin{multline}
	\chi (\lambda) = (a - \lambda)(d - \lambda) - bc 
	\\= (\lambda - \frac{a+d}{2})^2 + ad - bc - \frac{(a+d)^2}{4}
	\label{eq:charapol}
\end{multline}

Introducing the variables $\tau = \frac{a+d}{2}$ and $\eta = \frac{a-d}{2}$, we get the following condition for eigenvalue collapse,

\begin{equation}
		bc = -\eta^2.
		\label{eq:evcollapse}
\end{equation}

\begin{figure}
\centering
 \hspace{0in} \includegraphics[width=1.0\columnwidth]{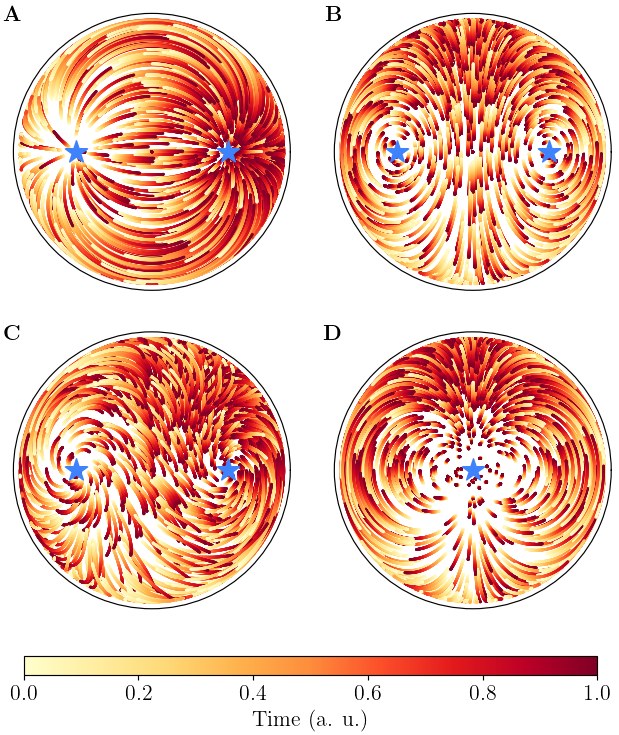}  \\
\caption{\textbf{The coalescence of polarisation.} Stereographic projection of the top hemisphere of the Poincaré sphere with 1000 random initial polarisations evolving under a Hamiltonian that correspond to (\textbf{A}) hyperbolic, (\textbf{B}) elliptic, (\textbf{C}) loxodromic, (\textbf{D}) parabolic M\"obius transformations, respectively. The polarisations of the eigenvectors are marked with blue stars.}
\label{fig1}
\end{figure}

Eigenvalue collapse is a necessary but insufficient condition for EPs to occur. Indeed, the case of an already diagonal $H$ with equal eigenvalues fullfills this condition, but possesses two independent eigenvectors. We leave out this case, since it leads to decoupled subsystems with trivial dynamics. Without loss of generality, we thus set $b\neq 0$. Introducing $\mu = c + \frac{\eta^2}{b}$, we obtain a convenient parametrization of non-diagonal complex $2\times 2$ Hamiltonians,
 
\begin{equation}
	H = \begin{pmatrix}
		\tau + \eta & b\\
		\mu-\frac{\eta^2}{b} & \tau - \eta
	\end{pmatrix},
	\label{eq:hdef}
\end{equation}

whose eigenvalues collapse if and only if $\mu = 0$. We now enforce this restriction, making $\tau$ the only eigenvalue of $H$. Proceeding towards the Jordan normal form, we define

\begin{equation}
	N := H - \tau \mathbb{I} = \begin{pmatrix}
		\eta & b\\
		-\frac{\eta^2}{b} & - \eta
	\end{pmatrix},
\end{equation}

which is nilpotent, as a direct computation shows. We can bring it to the canonical form using a similarity transform,

\begin{equation}
	 S^{-1}NS =
	\begin{pmatrix}
		\frac{1}{b} & 0\\
		\frac{\eta}{b} & 1
	\end{pmatrix}
    N
	\begin{pmatrix}
		b & 0\\
		-\eta & 1
	\end{pmatrix}
	=
	\begin{pmatrix}
		0 & 1\\
		0 & 0
	\end{pmatrix}
	.
\end{equation}

Applying the same transform to the Hamiltonian yields its Jordan normal form,

\begin{equation}
    S^{-1}HS =
	\begin{pmatrix}
		\tau & 1\\
		0 & \tau
	\end{pmatrix},
\end{equation}

where the single eigenvector associated to $\tau$ is given by 

\begin{equation}
	\ket{\phi} = \begin{pmatrix}b\\-\eta\end{pmatrix}
\end{equation}

in the original basis. The time evolution of $\ket{\phi}$ then obeys Eq.\ref{eq:HermSol}; the eigenstate simply undergoes damped or amplified oscillations and remains parallel to its initial state, exactly as eigenstates in the diagonalisable case. The key difference with Hermitian time-evolution is this: we can no longer express the time evolution of arbitrary states by projecting them on the eigenstates and evolve them separately, since they do not span the full Hilbert space. The remainder of the Hilbert space did not disappear, we simply cannot compute the time evolution in this subsector the usual way. 


\subsection{Time evolution at the EP}

To see what happens when the initial state is not fully aligned with the single eigenstate $\ket{\phi}$, we compute the evolution operator $U(t)$, a matrix exponential, by using the fact that the identity commutes with all other matrices. This allows us to use the Baker-Campbell-Hausdorff formula:

\begin{multline}
	U(t)=e^{-\frac{i t}{\hbar}H} = e^{-\frac{i t}{\hbar}(\tau \mathbb{I} + N)} = e^{-\frac{i t}{\hbar}\tau \mathbb{I}}e^{-\frac{i t}{\hbar}N}\\
	 = e^{-\frac{i t}{\hbar}\tau}\sum^{\infty}_{k=0}\frac{1}{k!}(-\frac{i t}{\hbar})^k N^k=e^{-\frac{i t}{\hbar}\tau}\sum^{n}_{k=0}\frac{1}{k!}(-\frac{i t}{\hbar})^k N^k,
	 \label{eq:EPcascade}
\end{multline} 

where $n$ is the order of the EP.  An implicit version of this result was presented in \cite{heiss_time_2010}. For the Hamiltonian of Eq.\ref{eq:hdef}, the sum is cut off at $n=1$, since $N^2=0$, but this formula is also valid for EPs of higher order. This yields the solution

\begin{equation}
	\ket{\psi(t)} =
	U(t) \ket{\psi(0)}=e^{\frac{-i\tau}{\hbar}t}(\mathbb{I}-\frac{it}{\hbar}N)\ket{\psi(0)}.
	\label{eq:ep_static_solution}
\end{equation}

On one hand, we see that applying Eq.\ref{eq:ep_static_solution} to the coalesced eigenstate $\ket{\phi}$ yields the expected time evolution, since $N\ket{\phi}=0$. On the other hand, states that are not proportional to $\ket{\phi}$ generate components along $\ket{\phi}$ linearly in time, as the image of $N$ is spanned by $\ket{\phi}$. Arbitrary states therefore get closer and closer to the eigenstate as time passes. Time evolution at EPs thus differs from generic Hamiltonians in the sense that only one state evolves with the usual oscillations (possibly with overall damping or amplification), while other states remain inextricably coupled to this single eigenstate. In the case of a higher-order EP (Eq.\ref{eq:EPcascade}), this translates into a cascading temporal evolution, where the generalized eigenvector of order $k$ (a vector in the kernel of $N^k$ but not of $N^{k-1}$) generates each generalized eigenvector of lower order according to a power law in time. Note that the highest power in time is associated to the generalized eigenvector of lowest order, namely the EP eigenstate. This cascading evolution remains decoupled from generalized eigenstates of order higher than $k$; in particular, the coalesced eigenstate is the only one with fully decoupled dynamics. The only way to escape the EP eigenstate dominance, as experimentally realized in \cite{chen_revealing_2020}, is to introduce a source, thus rendering Eq.\ref{eq:schroedinger} inhomogeneous.


\subsection{The four Möbius classes}

How different can the situation be at a point close to the EP? Setting $\mu$ to be nonzero again, we find that the characteristic polynomial becomes

\begin{equation}
	\chi (\lambda)  = (\lambda - \tau)^2 - b\mu,
\end{equation}

yielding the eigenvalues 

\begin{equation}
	\lambda  = \tau \pm \sqrt{b\mu}.
	\label{eq:riemann}
\end{equation}

Interpreted as a function of $\mu$, Eq.\ref{eq:riemann} defines the square root Riemann surface that is often seen as the hallmark of EPs. This Riemann surface appears naturally in the band structure of non-hermitian systems with periodic spatial extension, with a parametric dependency on the momentum and a gain-loss parameter playing the role of $\mu$ for a momentum-space Hamiltonian. Naturally, the splitting induced by $\mu$ also occurs at the eigenvector level, with $\ket{\phi}$ turning into a pair of eigenvectors,

\begin{equation}
	\ket{\phi}_{\mu}^{\pm} = \ket{\phi} \pm \begin{pmatrix}0\\ \sqrt{b\mu} \end{pmatrix}.
	\label{eq:EPmuVectors}
\end{equation}

In principle, we now have a complete eigenbasis with simple time evolution, on which we can project and evolve arbitrary states, as in the hermitian case. Nevertheless, the corresponding evolution operator relates continuously to the EP evolution operator. Indeed, an explicit computation yields

\begin{multline}
    U_{\mu}(t)=e^{-\frac{i t}{\hbar}H}\\=e^{\frac{-i\tau}{\hbar}t}\bigg[\cos\bigg(\frac{\sqrt{b\mu}}{\hbar}t\bigg)\mathbb{I}-\frac{it}{\hbar}\sinc\bigg(\frac{\sqrt{b\mu}}{\hbar}t\bigg)N\bigg],
    \label{eq:static_U_op}
\end{multline}

which reduces to the form given in Eq.\ref{eq:ep_static_solution} when $\mu\rightarrow0$, with the leading term of the difference being linear in $\mu$. To gain insight on the qualitative effect of this degeneracy lifting, we can start from the following observation: in general, nonzero values of $\sqrt{b\mu}$ will not only lead to a separation of eigenstates, but also dampen one of them and amplify the other, due to the imaginary part of the eigenvalue. In this situation, any fluctuation away from the decaying eigenstate will therefore be amplified and tend to align more and more with the dominating eigenstate. We can make this statement quantitative by investigating the temporal evolution of the state's polarisation. Defining the latter as $p(t):=\frac{\psi_2(t)}{\psi_1(t)}$, we get

\begin{multline}
    p(t)=\frac{\psi_2(t)}{\psi_1(t)}=\\ \frac{(\cos(\frac{\sqrt{b\mu}}{\hbar}t)+\frac{i\eta}{\sqrt{b\mu}}\sin(\frac{\sqrt{b\mu}}{\hbar}t))p_0+(\frac{i(\eta^2-b\mu)}{b\sqrt{b\mu}}\sin(\frac{\sqrt{b\mu}}{\hbar}t))}{(-\frac{ib}{\sqrt{b\mu}}\sin(\frac{\sqrt{b\mu}}{\hbar}t))p_0+(\cos(\frac{\sqrt{b\mu}}{\hbar}t)-\frac{i\eta}{\sqrt{b\mu}}\sin(\frac{\sqrt{b\mu}}{\hbar}t))},
    \label{eq:polar_moebius}
\end{multline}

which constitutes a time-dependent M\"obius transformation of the initial polarisation $p_0$. Such transformations are elements of the M\"obius group. After properly normalising the factors appearing in Eq.\ref{eq:polar_moebius}, we can use the square of the trace of the associated matrix \cite{Needham_1998} to classify the transformation: $\sigma:=\Tr^2(U/e^{\frac{-i\tau}{\hbar}t})=4\cos^2(\frac{\sqrt{b\mu}}{\hbar}t)$. The M\"obius group then splits into four classes:

\begin{itemize}

\item The \textit{hyperbolic} class (Fig.\ref{fig1}C) corresponds to $\sigma$ real with $\sigma>4$. For $b\mu$ real and negative, the transformation is hyperbolic at all times. In this case, all initial polarisations converge towards the attractive eigenstate along direct arcs, with the exception of an unstable initial condition corresponding to the repulsive eigenstate.

\item The \textit{elliptic} class (Fig.\ref{fig1}B) corresponds to $\sigma$ real with $\sigma<4$. For $b\mu$ real and positive, the transformation stays elliptic at all times, except for the periodic return to the initial polarisation. This case is the most hermitian-like: all polarisations oscillate in time and no eigenvector dominates.

\item The \textit{loxodromic} class (Fig.\ref{fig1}A) is the most generic one; it occurs for $\Im(\sigma)\neq0$. For complex $b\mu$, the transformation stays loxodromic almost all the time, except for isolated elliptic and hyperbolic points that occur alternately for $\Re(\frac{b\mu}{\hbar}t)= n\frac{\pi}{2}, n\in \mathbb{N}$. This class can be considered as an intermediary between elliptic and hyperbolic classes. As in the latter case, one eigenstate dominates, but the trajectories are now spiralling to some degree.

\item The \textit{parabolic} class (Fig.\ref{fig1}D) arises for non-identity transformations with $\sigma=4$. This occurs when the two fixed points of the other classes merge, which happens for $b\mu=0$; we then have an EP and the system stays parabolic at all times. All polarisations eventually converge to the unique fixed point corresponding to the coalesced eigenstate.

\end{itemize}

The value of $b\mu$ therefore largely determines the nature of the M\"obius transformation that acts on the polarisation. Hence, we will term the Hamiltonians themselves hyperbolic, elliptic, parabolic and loxodromic depending on the value of $b\mu$. Physically, since $b\mu$ is equal to the sum of the square of the detuning with the product of the hopping terms, the EP occurs when the couplings between the two resonators match the detuning with the correct phase and amplitude. If this condition is not met, the coalesced eigenstate can split in three different ways: elliptic, hyperbolic and loxodromic. Let us focus on the latter two, which correspond to the generic case of $\Im(\sqrt{b\mu})\neq 0$. Using $\lim_{t\rightarrow\infty} \tan(\frac{\sqrt{b\mu}}{\hbar} t)= \pm i$, we get

\begin{multline}
    \lim_{t\rightarrow\infty}p(t)\\=\lim_{t\rightarrow\infty}\frac{b\sqrt{b\mu}p_0+i(bp_0\eta+\eta^2-b\mu)\tan(\frac{\sqrt{b\mu}}{\hbar}t)}{b\sqrt{b\mu}-ib(bp_0+\eta)\tan(\frac{\sqrt{b\mu}}{\hbar}t)}\\=\frac{-\eta\pm\sqrt{b\mu}}{b},
    \label{eq:polarisation}
\end{multline}

which corresponds to the polarisation of the dominating eigenstate, namely the one with the eigenvalue of largest positive imaginary part. As in the parabolic (EP) case, the long-term fate of the polarisation does not depend on the initial condition for these generic non-hermitian Hamiltonians, unless the system is initialized exactly in the repulsive eigenstate of the system. For example, this means that generic non-hermitian waveguides will act as polarisers in the paraxial approximation.

\subsection{Pseudo-Hermiticity}

 What about the non-generic cases we set aside ? Starting from an EP, we introduce the deformation $\mu=s b^*$, with $s$ real. This defines a set that presents no loxodromic phase and in particular includes the elliptic cases (as $s>0$) which we had discarded to obtain Eq.\ref{eq:polarisation}. As for $s<0$, it corresponds to the hyperbolic class. Recalling Eq.\ref{eq:riemann}, we see that the spectrum is either purely real, in the elliptic case, or consists of a complex-conjugate pair, in the hyperbolic case. It was shown in \cite{mostafazadeh_pseudo-hermiticity_2002-1} that Hamiltonians have such spectra if and only if they commute with an invertible antilinear operator $A$,
 
 \begin{equation}
    [H,A]=0,
    \label{eq:Asym}
\end{equation}
 
which is also equivalent to the property of \textit{pseudo-hermiticity}, introduced in \cite{mostafazadeh_pseudo-hermiticity_2002}, which makes our above analogy between the elliptic and the hermitian case precise. While hermitian Hamiltonians are self-adjoint with respect to the canonical inner product of the Hilbert space, pseudo-hermitian Hamiltonians are self-adjoint with respect to a modified inner product. A well-known example of antilinear symmetry inspired this concept, namely $\mathcal{PT}$ symmetry \cite{bender_real_1998}. In the particular case of $A=\mathcal{PT}$, the hyperbolic and elliptic classes are respectively known as ``$\mathcal{PT}$-broken" and ``$\mathcal{PT}$-unbroken" phases. Such elliptic Hamiltonians are the exception that proves the rule: for most cases, we have shown that non-hermitian time evolution is dominated by a single eigenvector.



\section{Floquet non-Hermitian Hamiltonians}

\begin{figure}
\centering
 \hspace{0in} \includegraphics[width=1.0\columnwidth]{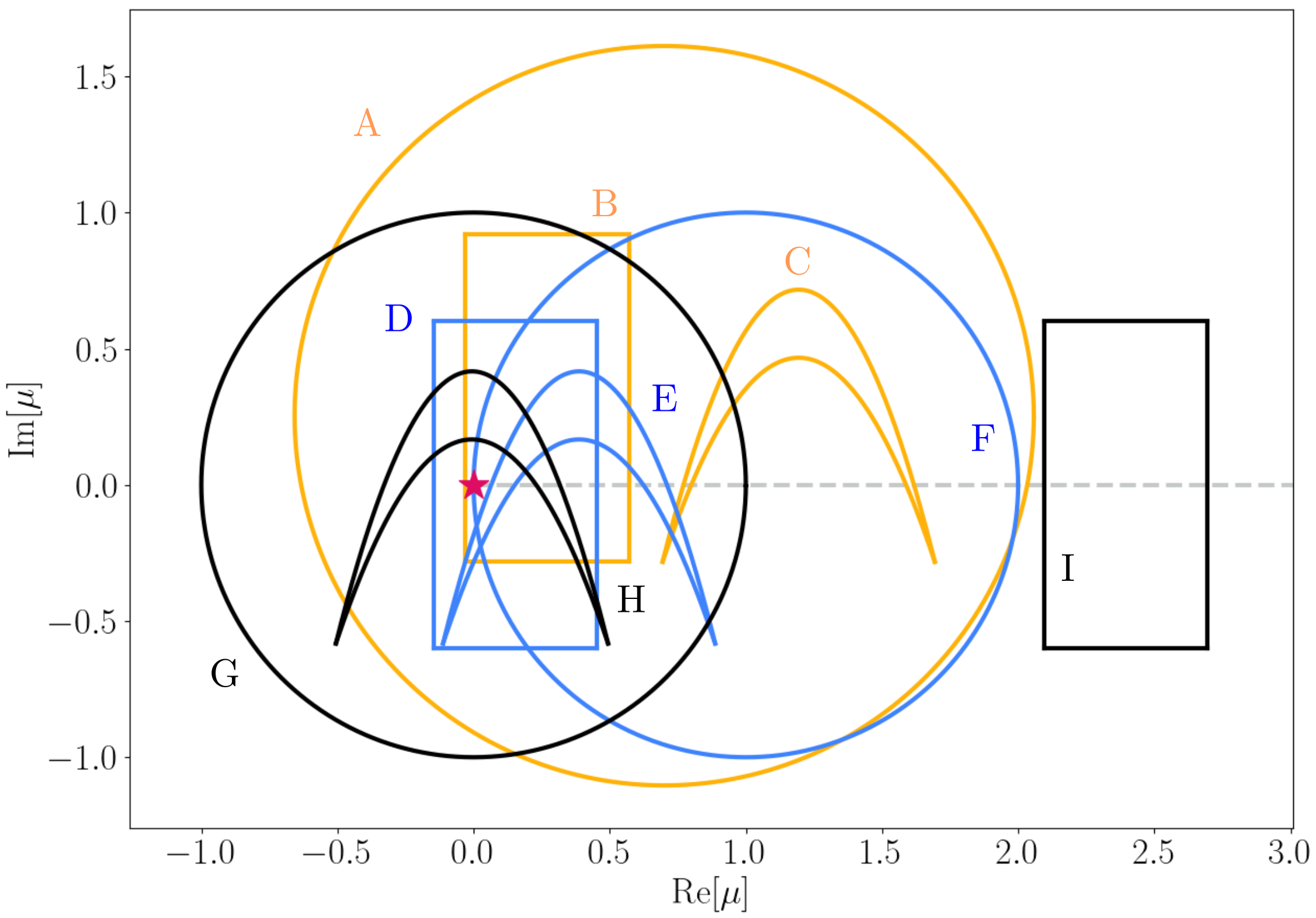}  \\
\caption{\textbf{Time modulation curves.} Modulation curves leading to loxodromic (A-C, yellow), elliptic (D-F, blue) and parabolic (G-I, black) time evolution. The time-independent EP is denoted by a red star. The grey dashed line corresponds to $Im(\sqrt{\mu})=0$.}
\label{fig:trajs}
\end{figure}

Varying the Hamiltonian's parameters in time in order to encircle the EP degeneracy has been theorized to produce geometric phase effects and asymmetric mode switching. In order to study this phenomenon, we derive exact solutions for time evolution under periodically-modulated Hamiltonians and interpret them in terms of our M\"obius group classification. The modulations we will study keep the EP Riemann surface fixed, with $\mu(t)$ and $\eta(t)$ describing some trajectories on the complex plane. Without loss of generality, we can set $\tau=0$, since it only gives rise to an exponential prefactor. We then rewrite equations \ref{eq:schroedinger} and \ref{eq:hdef} as

\begin{align} 
    i\hbar\partial_t\eta\psi_1 = i\hbar(\partial_t\eta)\psi_1&+\eta^2\psi_1+b\eta\psi_2
    \label{eq:psi2tied}\\ 
    i\hbar\partial_t b\psi_2 = b\mu\psi_1 &- \eta^2\psi_1 - b\eta\psi_2
\end{align}

which we sum to obtain

\begin{equation}
    p(t) \psi_{1} = -\hbar^2\partial_{t}^2\psi_{1},
    \label{eq:kode}
\end{equation}

where $p(t):=b\mu+i\hbar\partial_t\eta$. One can treat cases where $\tau$ and $b$ are also time-modulated in an analogous manner, but this will in general give rise to an additional damping term. For periodic modulations, Eq.\ref{eq:kode} is the Hill equation \cite{Hill_1886}, which has no general closed-form solutions. A particularly important sub-case is Mathieu's equation \cite{mathieu_equation_1868}, which corresponds to a purely real cosine modulation, giving rise to the phenomenon of parametric resonance \cite{strutt_beugung_1931}. Since our motivation lies in EP encircling, we consider modulations that describe loops in the complex plane instead, with segments for which Eq.\ref{eq:kode} has analytical solutions. For any such segment, we obtain two solutions $\psi_1$ and $\psi_1'$. As for $\psi_2$ and $\psi_2'$, they can be evaluated from $\psi_1$ through Eq.\ref{eq:psi2tied}. From these solutions, we then build a fundamental matrix,

\begin{equation}
    \Psi(t) := 
    \begin{pmatrix}
		\psi_1(t) & \psi_1'(t)\\
		\frac{i\hbar}{b}\partial_t\psi_1(t)-\frac{\eta}{b}\psi_1(t) & \frac{i\hbar}{b}\partial_t\psi_1'(t)-\frac{\eta}{b}\psi_1'(t)
	\end{pmatrix},
	\label{eq:PsiSol}
\end{equation}

from which the evolution operator follows,

\begin{equation}
    U(t)=\Psi(t)\Psi^{-1}(0).
\end{equation}

If the modulation changes after a time $t_c$ elapsed, the overall evolution operator becomes

\begin{equation}
    U(t)=(1-\Theta(t-t_c))U_a(t)+\Theta(t-t_c)U_b(t)U_a(t_c),
\end{equation}

where $U_a$ and $U_b$ are respectively the evolution operators associated to the first and second phase of the modulation. We can therefore form solutions for a large variety of piecewise defined modulation curves in the complex plane. In general, a given second-order differential equation can be realised by infinitely many different Hamiltonians, corresponding to different choices of $\mu$ and $\eta$. For the sake of simplicity, we will therefore set $\eta=0$ from now on, keeping in mind that $U$ can be deformed using Eqs.\ref{eq:kode} and \ref{eq:PsiSol} to accommodate for a time-dependent detuning $\eta$. 

We apply this method to three different families of periodically time-modulated Hamiltonians, with modulation curves that are constituted of linear (Fig.\ref{fig:trajs}A,F,G), circular (Fig.\ref{fig:trajs}B,D,I) and quadratic (Fig.\ref{fig:trajs}C,E,H) segments. The corresponding solutions to Eq.\ref{eq:kode} are respectively Airy, Bessel and Parabolic Cylinder functions \cite{NIST:DLMF}.

Floquet's theorem \cite{floquet_sur_1883} tells us that for such periodically modulated Hamiltonians, the evolution operator decomposes as $U(t)=Q(t)e^{Bt}$, where $Q(t)$ has the same periodicity as the modulation, and $B$ is a constant matrix. The long-term behaviour of the system is therefore determined by the non-periodic envelope $e^{Bt}$, which has the same form as the evolution operator for time-independent Hamiltonians. We can therefore also classify Floquet non-hermitian systems in terms of the M\"obius group. In the following, we employ this framework to show that the essence of time evolution in Floquet non-hermitian systems is not captured by EP encircling.

\subsection{M\"obius class versus encircling}

\begin{figure}
\centering
 \includegraphics[width=0.95\columnwidth]{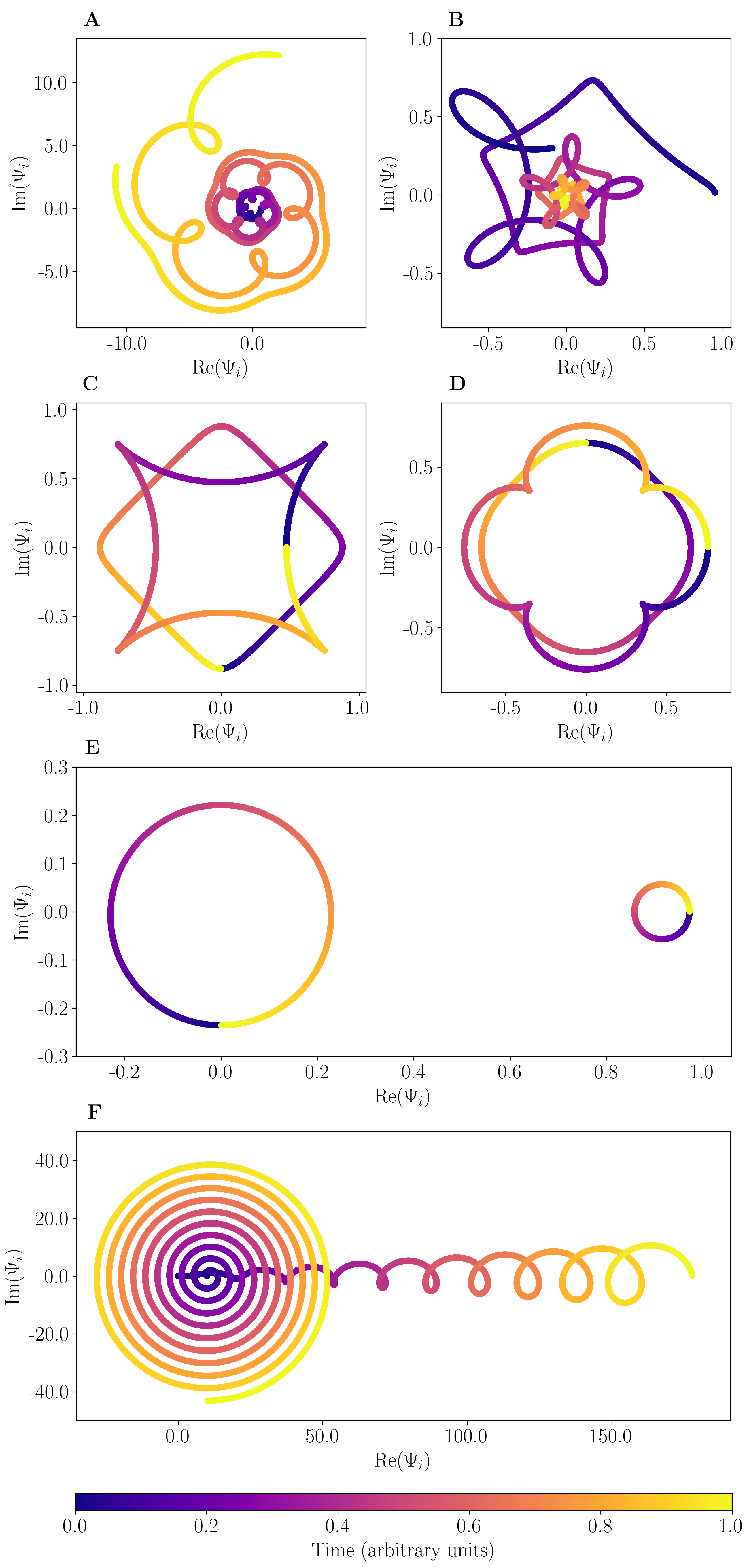} \\
\caption{\textbf{Circular modulations.} (\textbf{A-E}) Time evolution of stroboscopic eigenstates for various parameters. (\textbf{A-B}) , Non-centered EP encircling with $\Delta=0.700145+0.254176i$ and $\rho=1.357497$. (\textbf{C-D}) , Loop crossing the static EP, $\Delta=\rho=1$. (\textbf{E}) Centered EP encircling, with $\Delta=0$ and $\rho=1$. (\textbf{F}) Evolution of a non-eigenstate under the same parameters. The state's two components are represented in each plot, with time encoded as color.}
\label{fig:FEP1}
\end{figure}

We start with a modulation curve commonly considered in theoretical studies, namely a circular modulation (Fig.\ref{fig:trajs}AFG), which has the form 

\begin{equation}
    \mu(t)=\Delta+\rho e^{i\omega t}. 
\end{equation}

It can be solved in terms of Bessel functions, as in done in \cite{berry_slow_2011}. The Floquet eigenvalues evaluate to $\lambda=\sqrt{b\Delta}$, revealing that the time-dependent case is in close correspondence to the static one. Indeed, generic parameters lead to loxodromic evolution (Fig.\ref{fig:FEP1}A-B). In this case, the dominant eigenstate expands following a logarithmic spiral (Fig.\ref{fig:FEP1}A), while the other eigenstate spirals towards the origin (Fig.\ref{fig:FEP1}B). On the other hand, circles centered on the positive real axis lead to stable trajectories, which can be periodic (Fig.\ref{fig:FEP1}C-D) or form rosettes for irrational Floquet eigenvalues. In particular, periodic trajectories can have periods much larger than the modulation.

Last but not least, we get a Floquet EP (Fig.\ref{fig:trajs}G and Fig.\ref{fig:FEP1}E-F) when the modulation circle is centered on the static EP, regardless of the radius and modulation frequency. In Fig.\ref{fig:FEP1}E, we see that the eigenstate describes a cycle in the complex plane with every modulation period. In line with our earlier description of the time-independent case, Fig.\ref{fig:FEP1}F shows that other states tend to acquire the polarisation of the eigenstate, leading to a characteristic Archimedean spiral pattern. Crucially, circular modulations that encircle the EP without being exactly centered on it do not lead to a Floquet EP, since their Floquet eigenvalues are not degenerate.

The modulation depth $\rho$ does affect the periodic component of time evolution, $Q(t)$. For instance, if the modulation curve is a de-centered loop that dynamically crosses the static EP (Fig.\ref{fig:trajs}F), the evolution presents a cusp point at the time for which the EP condition is met (Fig.\ref{fig:FEP1}CD). For larger radii, the cusps turn into self-intersections (Fig.\ref{fig:FEP1}AB). However, this is also not a topological property, since self-intersections can be created in other ways.

\begin{figure}
\centering
 \includegraphics[width=0.95\columnwidth]{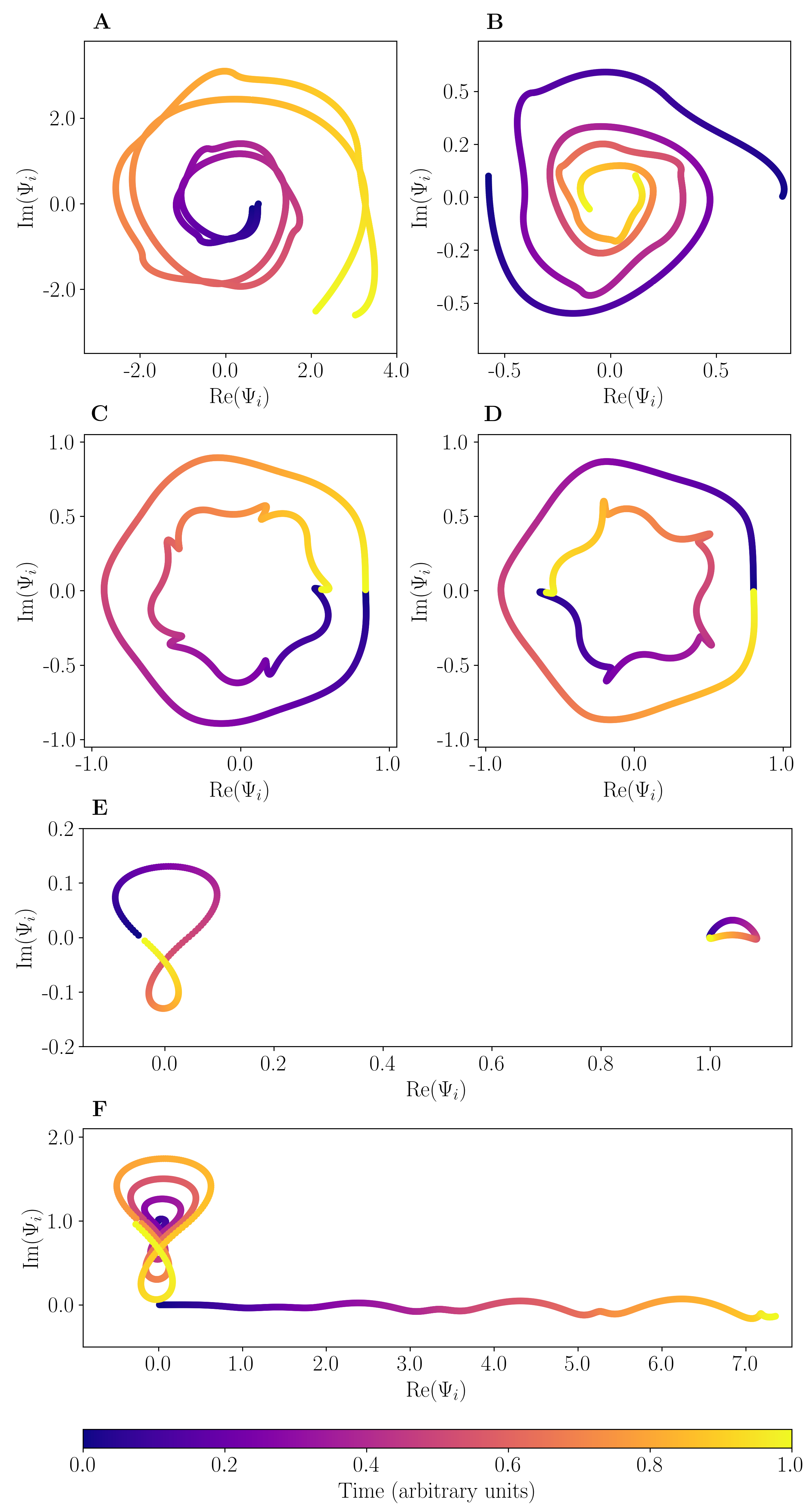} \\
\caption{\textbf{Quadratic modulations.} (\textbf{A-E}) Time evolution of stroboscopic eigenstates for various parameters. (\textbf{A-B}) Quadratic trajectory shifted by $\Delta=1.2+0.3i$. (\textbf{C-D}) Shifted by $\Delta=0.393399$. (\textbf{E}) Non-encircling $\Delta=0$ trajectory. (\textbf{F}) Evolution of a non-eigenstate under the same parameters. The state's two components are represented in each plot, with time encoded as color.}
\label{fig:FEP2}
\end{figure}


With these circular modulations, we have seen that encircling does not imply Floquet EP behaviour. We now show that Floquet EPs can also occur without encircling. To that end, we consider a concave modulation curve (Fig.\ref{fig:trajs}CEH) made of two quadratic segments,

\begin{align}
    \mu_a(t)&=\Delta-(1+i)/2 + (1 + 4i)t - 4it^2\\
    \mu_b(t)&=\Delta-(1+i)/2 + (1 + 3 i) (2 - t) - 3 i (2 - t)^2.
\end{align}

This case is solved in terms of parabolic cylinder functions. It also exhibits solutions of a loxodromic (Fig.\ref{fig:FEP2}A-B) and elliptic nature (Fig.\ref{fig:FEP2}C-D, special case of a rational eigenvalue). Without surprise, the periodic component of time-evolution is heavily impacted by the different modulation curve. More interestingly, we obtain a Floquet EP without encircling the static EP. At this Floquet EP, non-eigenstate initial conditions have an archimedean-spiral-like behaviour (Fig.\ref{fig:FEP2}F) that tends towards the eigenstate (Fig.\ref{fig:FEP2}E). The existence of a Floquet EP is therefore not tributary of the topological properties of the EP Riemann surface.


\begin{figure}
\centering
 \includegraphics[width=0.95\columnwidth]{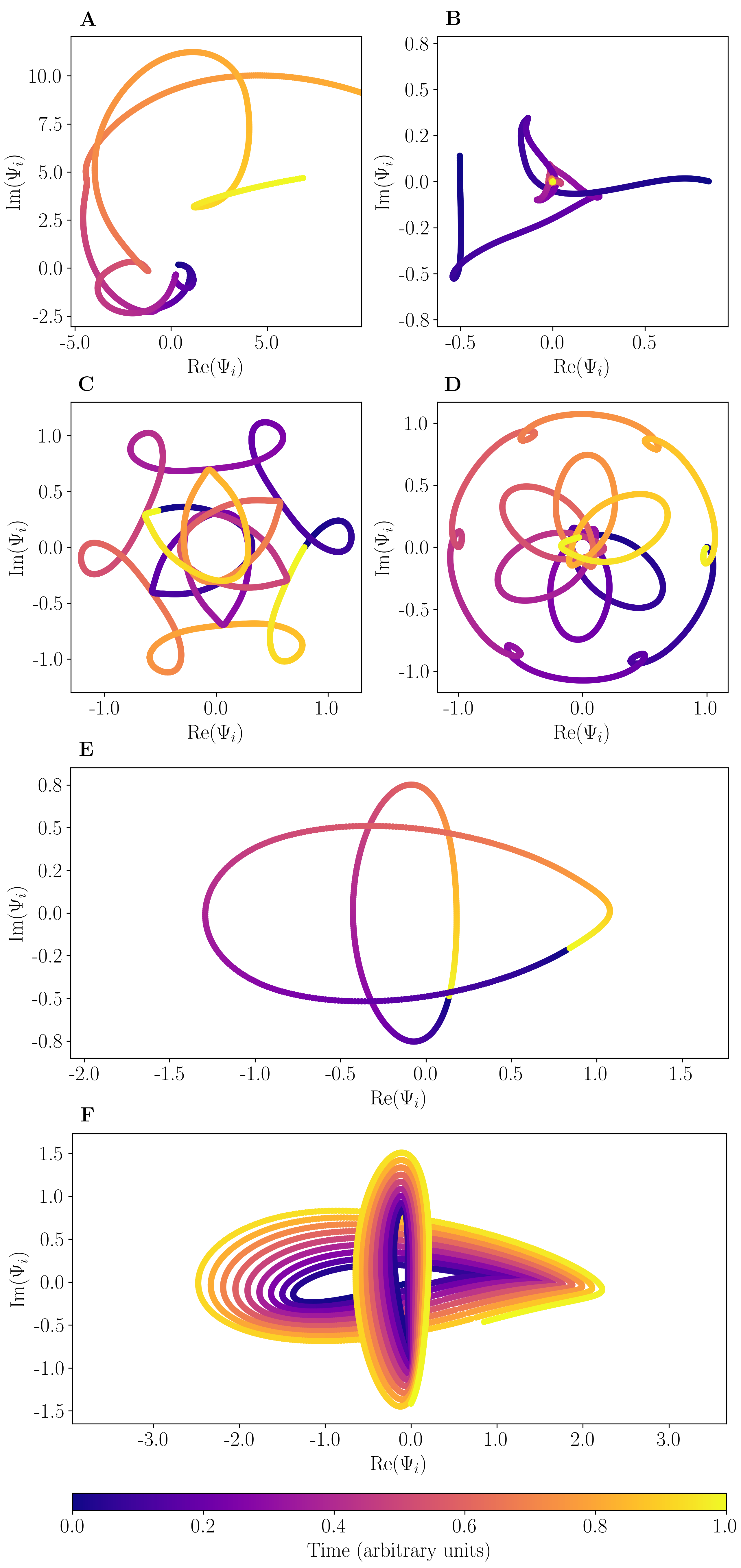} \\
\caption{\textbf{Rectangular modulations.} (\textbf{A-E}) Time evolution of stroboscopic eigenstates for various parameters. (\textbf{A-B}) Rectangular trajectory shifted by $\Delta=0.27+0.32i$. (\textbf{C-D}) Shifted by $\Delta=0.153$. (\textbf{E}) Distant $\Delta=2.394756696$ trajectory. (\textbf{F}) Evolution of a non-eigenstate under the same parameters. The state's two components are represented in each plot, with time encoded as color.}
\label{fig:FEP3}
\end{figure}


How do the M\"obius class and modulation relate to each other then ? Is there at least a direct relationship between the modulation center and the Floquet eigenvalues, as in the circular case? To answer this question, we consider a time-modulation with linear segments, which can be solved in terms of Airy functions. Our example comprises four such segments, defined by

\begin{align}
    \mu_a(t)&=\Delta - \frac{\rho}{2} + i\alpha\frac{\rho}{2} + \rho t\\
    \mu_b(t)&=\Delta + \frac{\rho}{2} + i\alpha\frac{\rho}{2} - i\alpha \rho t\\
    \mu_c(t)&=\Delta + \frac{\rho}{2} - i\alpha\frac{\rho}{2} - \rho t\\
    \mu_d(t)&=\Delta - \frac{\rho}{2} - i\alpha\frac{\rho}{2} + i\alpha \rho t,
\end{align}

which build a rectangle (Fig.\ref{fig:trajs}BDI) of aspect ratio $\alpha$ and width $\rho$. Again, we also find solutions of a loxodromic (Fig.\ref{fig:FEP2}A-B) and elliptic (Fig.\ref{fig:FEP2}C-D) nature. Again, the periodic component of time-evolution is different from the previous cases, and again, we find a Floquet EP without encircling the static EP (Fig.\ref{fig:FEP3}E-F). Furthermore, the time-modulation curve that gives rise to this particular Floquet EP lies far away from the static EP (Fig.\ref{fig:trajs}I), demonstrating that the behaviour observed in the circular case does not generalize. With these various examples, we have shown that the M\"obius class, which captures the nature of time evolution, is logically decoupled from EP encircling. In the next section, we show that such phenomena are better understood in terms of parametric resonance.


\begin{figure}
\centering
 \includegraphics[width=1.0\columnwidth]{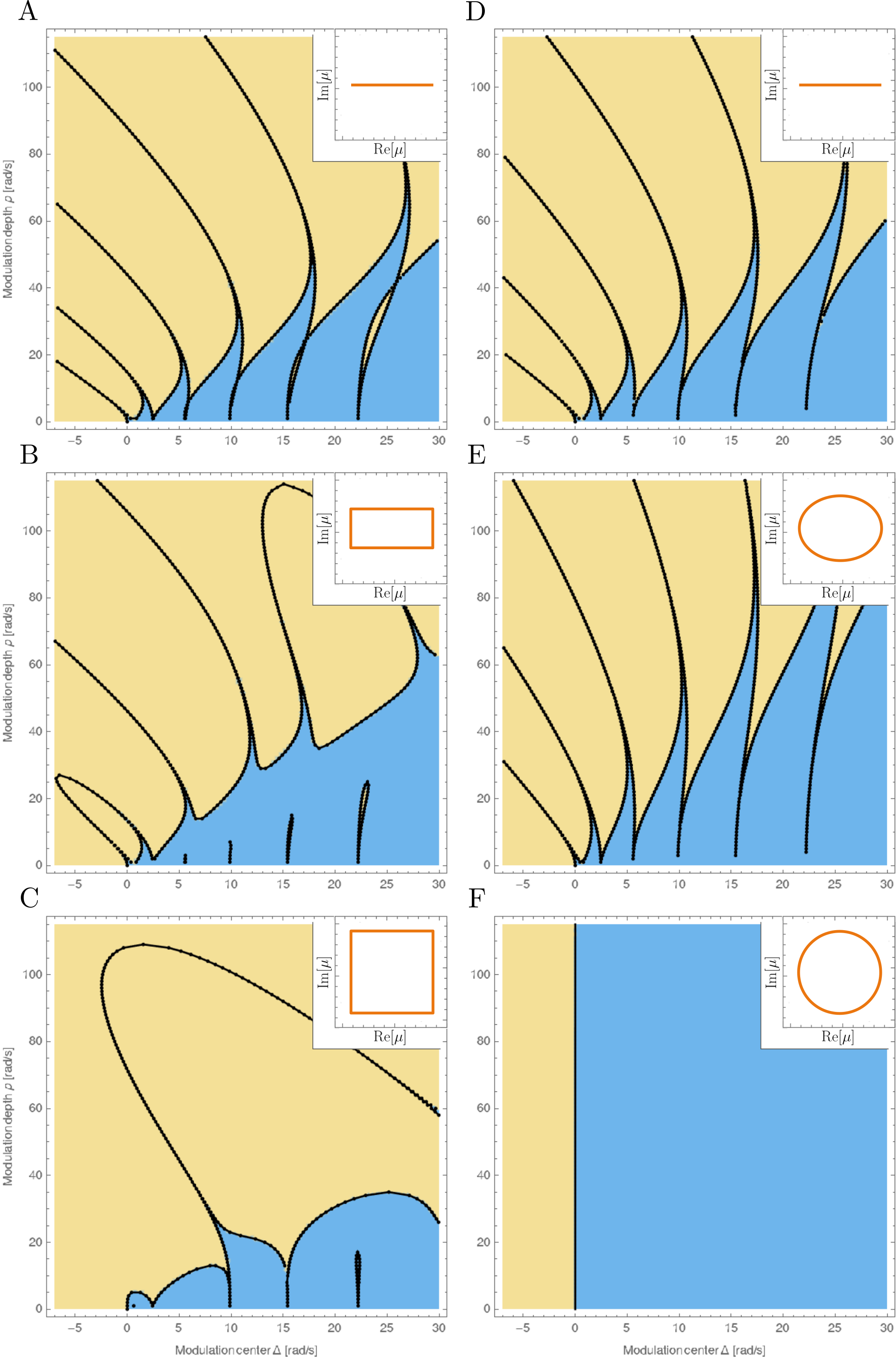} \\
\caption{\textbf{Non-hermitian parametric resonance} Stability diagrams for rectangular (\textbf{A-C}) and elliptic (\textbf{D-F}) modulations of varying aspect ratios. Black points denote EPs. The blue regions are in the stable (elliptic) phase, whereas the yellow regions are in the unstable (hyperbolic) phase. The aspect ratios are $\alpha=0$ (\textbf{A}), $\alpha=0.475$ (\textbf{B}), $\alpha=1.0$ (\textbf{C}), $\alpha=0.0$ (\textbf{D}), $\alpha=0.8$ (\textbf{E}), $\alpha=1.0$ (\textbf{F}). The corresponding modulation curves are represented in inserts.}
\label{fig:PR}
\end{figure}

\section{Non-hermitian parametric resonance}

The close correspondence between the static and circular Floquet case (Fig.\ref{fig:PR}F) is atypical; the other cases we studied hint at a much more intricate distribution of M\"obius classes, with Floquet EPs occuring far away from the static EP. We now study this distribution in more detail, by comparing stability diagrams for modulation curves describing rectangles (Fig.\ref{fig:PR}A-C) and ellipses (Fig.\ref{fig:PR}D-F) in the complex plane.

We start with a rectangular trajectory chosen to mimic Mathieu's equation, by setting the aspect ratio $\alpha=0$. The modulation is then purely real and approximates a cosine. The corresponding stability diagram (Fig.\ref{fig:PR}A) is qualitatively similar to the one obtained for a purely real cosine modulation (Fig.\ref{fig:PR}D), which is known as a Strutt diagram \cite{strutt_beugung_1931}. Both exhibit the essential features of parametric resonance, namely domains of stability (blue) and instability (yellow) that interpenetrate through several very thin "\textit{tongues}" of instability (resp. stability.), whose boundaries (black lines) correspond to Floquet EPs. Only one essential difference appears: in the rectangular case, the instability tongues all present a self-crossing of their two EP boundaries.

Let us instead try to imitate the circular case, by selecting a square time modulation with $\alpha=1$. As we can see in Fig.\ref{fig:PR}C, the result is entirely different from the circular one (Fig.\ref{fig:PR}F): the stability diagram still shows many non-trivial features, which now differ substantially from the ones observed in standard parametric resonance; the overall number of stability tongues has been reduced, and the remaining ones are no longer shooting off to infinity but instead form an arch, merging at some finite modulation depth. In stark contrast, the case of circular modulation shows no dependency on the modulation depth, with the stability regime corresponding exactly to the static case.

What about the intervening aspect ratios? They, too, are very different. In Fig.\ref{fig:PR}B, we show the case of a rectangular modulation with an aspect ratio of $\alpha=0.475$. No stability tongue has yet disappeared, but four of them (including the inverted pendulum one) have formed archs. Furthermore, the crossing points present in the instability tongues of Fig.\ref{fig:PR}A have now disappeared, splitting each instability tongue in an isolated domain within the now fully connected stable region and a part connecting to the larger instability domain. 

The effect of an intermediate aspect ratio is much simpler when the modulation follows an ellipse on the complex plane: Fig.\ref{fig:PR}E shows the stability diagram for an aspect ratio of $0.8$. The only effect is a stretching of the diagram that extends the reach of stability, ultimately culminating in the full correspondence observed in Fig.\ref{fig:PR}F. This reinforces the oddity of circular modulation curves: even a slight flattening of the circular trajectory introduces parametric instability. Given their atypicality, the use of circular trajectories in theoretical studies is rather unfortunate: for instance, one cannot hope to emulate a theoretical result based on circular modulation through an experiment employing a rectangular modulation, as the two differ greatly in their qualitative behaviour. Furthermore, even if circular loops were experimentally accessible, parametric resonance would creep back in at the slightest departure from circularity. Parametric resonance clearly needs to be accounted for in theoretical treatments of EP encircling.



\section{Concluding remarks}

First, we have shown how the peculiar cascading temporal dynamics of EPs relate continuously to that of neighbouring non-exceptional points. Classifying the possible temporal behaviours with the M\"obius group illuminates the position of EPs within the larger landscape of non-hermitian Hamiltonians and complements the usual approach based on restricting the parameter space through an antilinear symmetry.

This led to our main result; together with an analytical approach that highlights the connection to parametric resonance, these classes allowed us to also make sense of non-hermitian time-modulated systems. We demonstrated that such systems exhibit rich physics that cannot be reduced to the topological properties of the EP Riemannn surface; in particular, the occurrence of Floquet EPs is logically independent from whether the modulation curve encircles the static EP or not. Instead, we uncovered complex parametric resonance phenomena whose qualitative features greatly depend on the shape of the modulation curve. This creates exciting prospects, both in further exploration of the puzzling features of non-hermitian parametric resonance and in the possibility of synthesizing Floquet EPs through designed modulation schemes.

\section*{Acknowledgements}
A.B. thanks T. Koutserimpas for useful discussions.
A.B. and R.F. gratefully acknowledge funding from SNF grant 181232.

\section*{References}
\bibliography{References}


 


\end{document}